\documentclass[preprint,review,12pt,authoryear]{elsarticle}

\usepackage{lscape}
\usepackage{graphicx}
\usepackage[utf8]{inputenc}
\usepackage{amssymb}
\usepackage{amsmath}
\usepackage{natbib}
\usepackage[caption=false]{subfig}
\usepackage{subfig}
\usepackage{epstopdf}
\usepackage[normalem]{ulem}
\usepackage{lastpage}

\biboptions{}

\journal{International Journal of Solids and Structures}

\def\xR{\mathbb{R}}
\def\x0{\mathbf{0}}

\def\xi{\mathbf{i}}

\def\ds{\textrm{d}s}
\def\dt{\textrm{d}t}
\def\dtau{\textrm{d}\bar t}

\newtheorem{theorem}{Theorem}
\newtheorem{definition}{Definition}
\newtheorem{lemma}{Lemma}

\newtheorem{corollary}{Corollary}

\usepackage{xcolor}

\begin{document}
\begin{frontmatter}



\title{Shape of an elastica under growth restricted by friction}


\author[label1]{Marcell G. Horv\'ath}
\author[label1,label2]{Andr\'as A. Sipos}
\author[label1]{P\'eter L. V\'arkonyi}

\address[label1]{Department of Mechanics, Materials and Structures, Budapest University of Technology and Economics, Műegyetem rkp. 1-3. K261, Budapest 1111, Hungary}

\address[label2]{MTA-BME Morphodynamics Research Group, Budapest, Hungary}

\begin{abstract}
We investigate the quasi-static growth of elastic fibers in the presence of dry or viscous friction. An unusual form of destabilization beyond a critical length is described. In order to characterize this phenomenon, a new definition of stability against infinitesimal perturbations over finite time intervals is proposed and a semi-analytical method for the determination of the critical length is developed. The post-critical behavior of the system is studied by using an appropriate numerical scheme based on variational methods. We find post-critical shapes for uniformly distributed as well as for concentrated growth and demonstrate convergence to a figure-8 shape for large lengths when self-crossing is allowed. Comparison with simple physical experiments yields reasonable accuracy of the theoretical predictions.
\end{abstract}


\end{frontmatter}


\section{Introduction}

\label{Intro}
The loss of stability of slender elastic structures under external loads has been studied for centuries. In classical problems of elastic stability (like Euler buckling), the system is subject to conservative loads and stability of a static solution against infinitesimally small perturbations depends on second variations of an energy functional. Non-conservative forces like a follower load \citep{langthjem2000dynamic}, magnetic forces \citep{sinden2009spatial} or sliding friction  can be treated either via an extended energy approach or by direct analysis of the equations of motion. The loss of elastic stability plays a key role  in many processes related to biological growth, which is subject of the theory of morpho-elasticity \citep{goriely2017mathematics}. We contribute to this theory by investigating the stability of straight configurations of an elastic rod, which expands quasi-statically, while its growth is resisted by friction. 

In classical problems of elastic stability, destabilization occurs as some parameter (e.g. load intensity or friction coefficient) is slowly varied and crosses a critical value. This situation can be modeled as an autonomous slow-fast dynamical system, and stability analysis focuses on qualitative properties of the fast dynamics of the system as fast time goes to infinity. The problem investigated in the present paper has a fundamentally different character. Here, the expansion of the rod plays the role of the varying parameter, nevertheless the same expansion drives the quasi-static motion of the rod, which would stop in the absence of growth due to friction. Hence, the two  processes occur on the same time scale. Thereby we need to investigate a general non-autonomous dynamical system without time scale separation.

We will focus on those situations, when the thickness of the growing rod is constant over time, which makes it more and more susceptible to loss of stability due to frictional forces. Hence, we would like to characterize systems, which- loosely speaking - remain stable for a finite amount of time and become unstable thereafter. Unfortunately, this type of behaviour is somewhat ill-defined mathematically, since standard notions of stability (such as Lyapunov or asymptotic stability) are not applicable to finite time-intervals: stability inherently requires appropriate system behaviour as $t\rightarrow\infty$ (e.g. convergence to the unperturbed solution in the case of asymptotic stability). 

The notion of 'Finite-time stability'  \citep{amato2014finite} is frequently used in engineering control problems. This definition requires that for initial conditions within an $\epsilon$ neighborhood of the examined solution, the system stays withing some $\delta$ neighborhood of that solution over the entire finite time interval. 
Here - unlike in the case of classical Lyapunov stability of infinite time systems -, $\epsilon$ and $\delta$ are a-priori chosen finite scalars. This is not appropriate to us since this definition cannot deal with infinitesimally small perturbations. To avoid this caveat, we will develop a new notion of finite-time stability, which is closely related to the classical notion of exponential (or linear) stability, but is applicable to finite time-intervals.
Using the newly developed concept of \emph{finite-time exponential stability}, we identify semi-analytically for some simple loads and constraints the critical length of the rod, where the straight configuration becomes unstable. We also point out fundamental differences between this phenomenon and more common destabilization phenomena like buckling and fluttering: instead of an abrupt change in the behavior of the rod, it undergoes slow divergence from the straight configuration, which becomes visible only after some additional growth.

The post-critical behavior of the system is strongly nonlinear and thus it is studied numerically. We find that there is no well-defined "buckled" shape: the rod continuously changes its shape after reaching the critical length, which is not surprising in a non-autonomous system. As growth continues, some parts of the rod undergo slip-stick transition if friction includes a Coulomb-type component, and the shape of the rod forms a figure-8 pattern. 

Finally, we also present experimental results, which show good agreement with numerical predictions.

\section{Model development}
\label{Sect2}

\subsection{Assumptions and governing equations}
We consider a planar, growing elastic rod (assumed to follow the Bernoulli-Euler beam model) where friction is acting between the rod and the plane as it grows. We formulate the governing equations as the first variation of the generalized total stored energy of the system. Energy consist of the strain energy ($E$) of the elastic rod and the dissipation energy ($D$) associated with friction.

Let $[0,a]$ denote a closed interval in $\xR$. The time-dependent shape of the rod is represented by a parametrized planar curve with a parameter $s\in [0,1]$. Between the time instances $t=0$ and $t=T$ the shape is given by the function $x: [0,1]\times [0,T]\rightarrow\xR^2$. We assume sufficient smoothness of $x$, in particular $x(s,t)\in C^4([0,1])\times C([0,T])$. To simplify notation $(.)'$ and $\dot{(.)}$ stand for the partial derivatives respect to the spatial and temporal variable, respectively. The size of the growing curve at a given time instant is represented by the $v(s,t)$ 'speed' \citep{DGbook} or 'growth-induced stretch' \citep{goriely2017mathematics} of the curve, where $v: [0,1]\times [0,T]\rightarrow\xR$ is formally defined as
\begin{equation}
\label{eq:growth}
v(s,t):=\left\|x'(s,t)\right\|
\end{equation}
\noindent with $\left\|.\right\|$ denoting the Euclidean norm. The arc length of the curve between $0$ and $s$ at time $t$ is denoted by
\begin{equation}
p(s,t):=\int_0^s v(\varphi,t)\textrm{d}\varphi.
\end{equation}

Note that natural parametrization of the curve during the evolution would mean $v(s,t)\equiv 1$ and $p(s,t)\equiv s$. 

We assume that the rate of growth is \emph{a-priori} known and thus the stretch function takes prescribed, sufficiently smooth values 
\begin{equation}
\label{eq:constr}
v(s,t)
\equiv \hat{g}(s,t).
\end{equation}

Assuming a straight stress-free configuration for the planar rod with a constant bending stiffness $YI$, the strain energy is formulated as
\begin{equation}
E(t)=\frac{YI}{2}\int_0^1\kappa(s,t)^2v(s,t) \ds,
\end{equation}

\noindent where $\kappa(s,t)$ is the curvature. $v(s,t)$ appears in $E$, because during the evolution we do not have natural parametrization for the curve. At time $t$, let $q(s,t)$  and $n(s,t)$ denote the unit tangent and normal vectors of the curve, respectively. The Frenet-Serret formulas deliver
\begin{equation}
\label{FS formulas}
x''(s,t)=v'(s,t)q(s,t)+v(s,t)^2\kappa(s,t)n(s,t).
\end{equation}
Taking dot products of each sides in eq. (\ref{FS formulas}), using the orthogonality of the Frenet-frame and applying algebraic manipulations lead to
\begin{equation}
\label{eq:kappa}
\kappa(s,t)^2=\frac{x''(s,t)\cdot x''(s,t)-v'(s)^2}{v(s,t)^4}=\frac{\left\|x''(s,t)\right\|^2-v'(s,t)^2}{v(s,t)^4}.
\end{equation}

About the dissipation energy we assume, that there is a distributed frictional force between the curve and its supporting plane, which is assumed to be some linear combination of Coulomb and linear viscous dissipations with constant coefficients $\mu$ and $\nu$:
\begin{equation}
f(s,t)=-v(s,t)\dot x(s,t)\left(\nu+\mu \frac{1}{\left\|\dot x(s,t)\right\|}\right).
\end{equation}
Note, that space and time dependent variation of the friction coefficients are natural extensions of our model. The energy density $\Psi$ associated with the infinitesimal segment of the rod between time instances $t_1$ and $t_1+\dt$ is given by
\begin{equation}
\label{eq:friction_ed}
\Psi(s,t_1):=\mu\int_{t_1}^{t_1+\dt}\left\|\dot{x}(s,\tau)\right\|d\tau+\frac{\nu}{2}\int_{t_1}^{t_1+\dt}\left\|\dot{x}(s,\tau)\right\|^2 \dtau.
\end{equation}

\noindent The Taylor-expansion of $x(s,t)$ with respect to time, at $t=t_1$, reads
\begin{equation}
x(s,t_1+\dt)=x(s,t_1)+\dot{x}(s,t_1)\dt+\mathcal{O}(\dt^2).
\end{equation}

\noindent Neglecting the nonlinear terms renders eq.(\ref{eq:friction_ed}) into

\begin{eqnarray}
\nonumber \Psi(s,t_1)=\mu\left\|x(s,t_1+\dt)-x(s,t_1)\right\|+\frac{\nu\left\|x(s,t_1+\dt)-x(s,t_1)\right\|^2}{2\dt}=\\
=\mu\left\|\dot{x}(s,t_1)\right\|\dt+\frac{\nu}{2}\left\|\dot{x}(s,t_1)\right\|^2\dt.
\end{eqnarray}

\noindent We drop the subscript in $t_1$ and deduce that the $D(t)$ dissipation energy whole along the rod is 
\begin{align}
D(t)&=\int_{0}^{1}\Psi(s,t)v(s,t)\ds   \\
&\approx\int_{0}^{1}\left\{\mu\sqrt{\left\|\dot{x}(s,t)\right\|^2+\beta}\dt+\frac{\nu}{2}\left\|\dot{x}(s,t)\right\|^2\dt \right\}v(s,t)\ds,
\end{align}

\noindent where in accordance with \citep{Capatina} we apply the small, fixed constant $\beta$ to regularize the functional and thereby establish the stability of the numerical solver. $\beta=0$ returns the exact value of $D(t)$.

We aim a \emph{quasi-static} description, in other words, we investigate a sequence of equilibrium configurations of the rod at a fixed $\dt$. At an equilibrium configuration the first variation of the total energy vanishes. Based on the above derivation and assuming $x(s,t-\dt)$ is known, the functionals $E$ and $D$ can be expressed as a function of $x(s,t)$, however, they both also depend on $v(s,t)$. As the growth function is assumed to be \emph{a-priori} known, a Lagrange-multiplier field $\lambda(s,t)$ is introduced to enforce \eqref{eq:constr} at each point and each time-step. The Lagrangian associated with the problem is formulated as

\begin{equation}
L(t):=E(t)+D(t)-\int_0^1\lambda(s,t)\left(\left\|x'(s,t)\right\|^2-\hat{g}^2(s,t)\right)\ds.
\end{equation}

Let $\eta(s)$ and $\zeta(s)$ denote admissible variations of $x(s,t)$ and $\lambda(s,t)$, respectively, at a fixed $t$. The first variation of the Lagrangian delivers the weak form of the governing equations as

\begin{eqnarray}
\label{GE:weak form}
\nonumber\delta L(t)=\int_{0}^{1}\left\{\frac{YI}{\hat{g}^3(s,t)}x''(s,t)\cdot\eta''(s)+\right.\\
\nonumber\left.\left(\nu+\frac{\mu}{\sqrt{\left\|\dot{x}(s,t)\right\|^2+\beta}}\right)\hat{g}(s,t)\dot{x}(s,t)\cdot\eta(s)-\right.\\
\left.-2\lambda(s,t)x'(s,t)\cdot\eta'(s)-\left(\left\|x'(s,t)\right\|^2-\hat{g}^2(s,t)\right)\zeta(s)\right\}\ds=0.
\end{eqnarray}

\subsection{Nondimensionalization}
\label{Sec:NonDim}

In order to identify key parameters, we develop a dimensionless form of our equations. Let
\begin{align}
V&:=\int_0^1\dot{\hat{g}}(s,0)\ds,\\
N&:=\mu+\nu V,
\end{align}

\noindent where $V$ represents the time derivative of the  arclength 
of the rod at $t=0$ and $N$ is a reference value of frictional forces. With these in hands we can introduce the following dimensionless variables:
\begin{align}
\bar t&=tV(YI)^{-1/3}N^{1/3},\\
\bar L&=L(YI)^{-2/3}N^{-1/3},\\
\bar g&=\hat{g}(YI)^{-1/3}N^{1/3}, \bar p=p(YI)^{-1/3}N^{1/3},\\
\bar x&=x(YI)^{-1/3}N^{1/3},\bar\eta=\eta(YI)^{-1/3}N^{1/3},\label{eq:lengthscaling}\\
\bar\mu&=\mu N^{-1},
\bar\lambda=\lambda N^{-1},\bar\zeta=\zeta N^{-1},\\
\bar\beta&=\beta V^{-2},
\end{align}

\noindent rendering the weak form of the governing equations into
\begin{eqnarray}
\label{GE:weak nondym}
\nonumber\delta \bar{L}(\bar{t})=\int_{0}^{1}\left\{\frac{1}{\bar{g}^3(s,\bar{t})}\bar{x}''(s,\bar{t})\cdot\bar{\eta}''(s)+\right.\\
\nonumber\left.\left(1-\bar\mu+\frac{\bar\mu}{\sqrt{\left\|\mathring{\bar{x}}(s,\bar{t})\right\|^2+\bar{\beta}}}\right)\bar{g}(s,\bar{t})\mathring{\bar{x}}(s,\bar{t})\bar\eta(s)-\right.\\
\left.-2\bar\lambda(s,\bar{t})\bar{x}'(s,\bar{t})\bar\eta'(s)-\left(\left\|\bar{x}'(s,\bar{t})\right\|^2-\bar{g}^2(s,\bar{t})\right)\bar{\zeta}(s)\right\}\ds=0.
\end{eqnarray}

Here, circle refers to the derivation with respect to rescaled time $\bar{t}$. The dimensionless form shows that the spatial distribution of growth does matter but its overall rate does not; that the values of $\nu$ and $\mu$ do not influence the shape of the growing rod as long as their ratio is constant; and finally that the absolute value of $YI$ is not important but the ratios $\mu/(YI),\nu/(YI)$ are. Note also that $\bar{\mu}=0$ corresponds to pure viscous friction and $\bar{\mu}=1$ to pure dry friction and that $\bar{g}$ is subject to the constraint
\begin{equation}
\int_0^1\mathring{\bar{g}}(s,0)\ds=1.
\label{eq:qconstraint}
\end{equation}

The Euler-Lagrange equations can be determined from the weak form by partial-integration. We will present simple examples below and note that the choice of $\bar{g}(s,\bar{t})$ may result in a rather complicated system of nonlinear PDEs.

As we solve a boundary value problem in each time-step, we need to clarify the boundary conditions. We assume, that the rod is clamped at $s=0$ and it is free at $s=1$. This latest is associated with vanishing internal moment and shear at $s=1$, so our simple linear elastic constitutive law yields $\bar{\kappa}(1,\bar{t})=\bar{\kappa}'(1,\bar{t})=0$, where $\bar{\kappa}$ is dimensionless curvature obtained as
\begin{equation}
\bar\kappa=\kappa(YI)^{1/3}N^{-1/3}.
\end{equation}

Applying the non-dimensionalized form of eq. (\ref{eq:kappa}) the boundary conditions are found to be
\begin{align}
\label{GE:BC1}
\bar{x}(0,\bar{t})=(0,0), &\qquad \bar{x}'(0,\bar{t})=(\bar{g}(0,\bar{t}),0), \\
\left\|\bar{x}''(1,\bar{t}) \right\|=\bar{g}'(1,\bar{t}), & \qquad \bar{x}''(1,\bar{t})\cdot\bar{x}'''(1,\bar{t})=\bar{g}'(1,\bar{t})\bar{g}''(1,\bar{t}).
\label{GE:BC2}
\end{align}

In the following we analyze two simple cases: a rod under uniform growth and an approximation of a rod growing at its fixed end. Note, that the trivial solution can be obtained for any continuous growth function as $\bar x_2\equiv 0$ where the notation $\bar{x}=[\bar{x}_1,\bar{x}_2]^T$ has been used. Then, $\bar x_1$ is determined uniquely by the non-dimensional form of the constraint equation \eqref{eq:constr}
\begin{equation}
\label{eq:constr2}
\left\| \bar x'(s,\bar t) \right\|
\equiv \bar{g}(s,\bar t)
\end{equation}
as the integral of the prescribed stretch function $\bar{g}(s,\bar t)$ with respect to $s$.

\subsection{Linear uniform growth}
In this subsection we restrict ourselves to uniform linear growth with spatial independence. In general, its growth function is $\bar{g}(s,\bar{t})=\bar{g}(0,\bar{t})+\bar{t}$. Without restricting generality, natural parametrization of the curve can be assumed at $\bar t=0$, hence
\begin{equation}
\label{eq:ugrowth}
\bar{g}(s,\bar{t})=1+\bar{t}.
\end{equation}

The Euler-Lagrange equations are derived from eq. (\ref{GE:weak nondym}) via integration by parts:
\begin{eqnarray}
\label{GE:strong form}
\frac{1}{(1+\bar{t})^3}\bar{x}_1''''+\left(\frac{\bar{\mu}}{\sqrt{\left\|\mathring{\bar{x}}\right\|^2+\bar\beta}}+1-\bar{\mu}\right)(1+\bar{t})\mathring{\bar{x}}_1+2\bar{\lambda} \bar{x}_1''+2\bar{\lambda}'\bar{x}_1'=0,\\
\label{GE:strong form1}
\frac{1}{(1+\bar{t})^3}\bar{x}_2''''+\left(\frac{\bar{\mu}}{\sqrt{\left\|\mathring{\bar{x}}\right\|^2+\bar\beta}}+1-\bar{\mu}\right)(1+\bar{t})\mathring{\bar{x}}_2+2\bar{\lambda} \bar{x}_2''+2\bar{\lambda}'\bar{x}_2'=0,\\
\label{GE:strong form2}
\left\|\bar{x}'\right\|^2-(1+\bar{t})^2=0,
\end{eqnarray}
The trivial solution of the system of governing equations is
\begin{eqnarray}
\label{GE:trivial}
\bar{x}(s,\bar{t})=[s(1+\bar{t}),0]^T,\\
\bar{\lambda}(s,\bar{t})=-\bar{\mu}\frac{\sqrt{s^2+\bar{\beta}}}{2}-(1-\bar{\mu})\frac{s^2}{4}.
\end{eqnarray}

\noindent where $\bar{x}$ is found from eqs. (\ref{GE:strong form1}) and (\ref{GE:strong form2}), whereas $\bar{\lambda}$ is expressed from (\ref{GE:strong form}). Linearization around the trivial solution delivers 
\begin{eqnarray}
\label{GE:linear}
\nonumber\left(\frac{\bar{\mu}}{\sigma}+1-\bar{\mu}\right)(1+\bar{t})\mathring{\bar{x}}_2=\\
-\frac{1}{(1+\bar{t})^3}\bar{x}_2''''+\left(\bar{\mu}\sigma+\frac{1}{2}(1-\bar{\mu})s^2\right)\bar{x}_2''+
\left(\frac{\bar{\mu} s}{\sigma}+(1-\bar{\mu})s\right)\bar{x}_2'
\end{eqnarray}

\noindent with $\sigma=\sqrt{s^2+\bar{\beta}}$.

\subsection{Growth concentrated at the end}
In the case of growth concentrated at the clamped end ($s=0$) we either choose to prescribe $\bar{g}(s,\bar{t})$ with a distribution (i.e. a Dirac-delta) or we take its continuous approximation. As either case leads to cumbersome expressions, and we perform a numerical simulation here, a continuous approximation (a bump function) is sufficient. Keeping the initial shape being naturally parametrized at $\bar t=0$, concentrated growth can be well approximated by

\begin{equation}
\label{eq:cgrowth}
\bar{g}(s,\bar{t})=1+b\exp\left(-\frac{s^2}{a^2}\right)\bar{t},
\end{equation}

\noindent with a fixed constant $a$. Parameter $b$ is determined uniquely by the constraint equation \eqref{eq:qconstraint}:
\begin{equation}
\label{eq:b}
b=\frac{2}{a\sqrt{\pi}\text{erf}{\frac{1}{a}}},
\end{equation}

\noindent where $\text{erf}(.)$ stands for the Gaussian error function. The trivial solution for this kind of growth is found to fulfill
\begin{equation}
\label{GE:trivial2}
\bar{x}(s,\bar{t})=[s+\frac{1}{2}ab\cdot\text{erf}\left(\frac{s}{a}\right) \bar{t},0]^T
\end{equation}
\noindent followed by a nasty expression in $\bar\lambda(s,\bar t)$. 

The growth function in eq. (\ref{eq:cgrowth}) realizes a transition between the continuous and concentrated growth as $a$ is varied: in the $a\rightarrow \infty$ limit this model exhibits uniform growth. On the other hand, $a\rightarrow 0$ concentrates the growth at the endpoint $s=0$ with no growth for any $s>0$.

\section{Model predictions}
\label{Sec:num}

\subsection{Numerics}
The finite element discretization of the weak form in equation (\ref{GE:weak form}) accompanied by the boundary conditions in eqs. (\ref{GE:BC1}) and (\ref{GE:BC2}) was implemented in FEniCS 1.6.0 \citep{fenics}. The quasi-static approach is reflected in the numerics: we seek equilibrium of the system for fixed $\bar t$, hence finite-element discretization is needed only in the space $s$. As the time-derivatives in our model are approximated by a forward Euler method, the time steps should be limited. In all of our computations $\Delta \bar{t}=0.01$ were applied and it granted convergence. We used an equidistant mesh for the spatial discretization of the unit interval $[0,1]$ with $N=500$ finite elements. We used a mixed finite-element space for the functions $\bar x_1(s,.)$, $\bar x_2(s,.)$ and $\bar\lambda(s,.)$ with a degree 3 polynomial approximation for each. As we treat a system of fourth-order PDEs, the interior penalty method was applied along the element boundaries \citep{Cockburn}. In all of our computations we fixed $\bar{\beta}=10^{-3}$ and the relative tolerance of the Newton solver at $\text{tol}=10^{-4}$. The initial length to start the simulations was $\bar p(1,0)=0.5$ in all cases.

\subsubsection{Uniform growth}
\begin{figure}[!ht]
\centering
\includegraphics[width=0.95\textwidth]{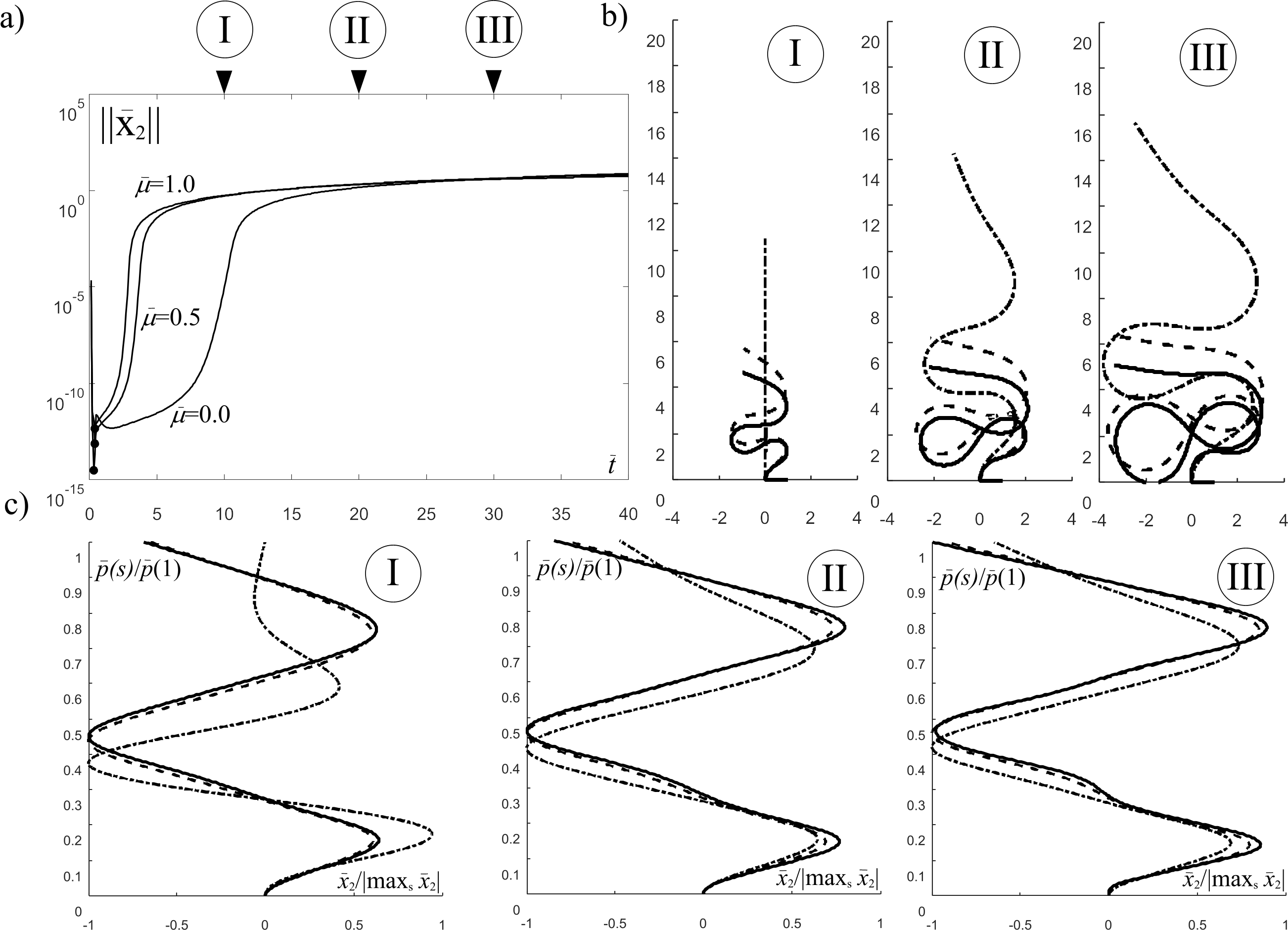}
\caption{Uniform growth at $\bar\mu=1.00$ (solid line), $\bar\mu=0.5$ (dash line) and $\bar\mu=0.0$ (dash-dot line). (a) Norm of the difference between the computed and the trivial solutions in $x_2$. Black dot denotes the minimum of that difference. (b) Physical realization of the computed curves at $\bar{t}=10.0, 20.0, 30.0$. (c) Normed shapes for the same curves.}
\label{Fig:01}
\end{figure}

In our non-dimensionalized model the response of the rod under the variation of the parameter $\bar{\mu}$ is a key question. We compute three cases, namely $\bar{\mu}=1.0,0.5$, and $0.0$. The evolution is started from some random shape obtained by adding a small perturbation to the trivial solution and it is followed up to $\bar t=40$ (altogether 4000 timesteps). The norm of the difference between the actual value of $\bar{x}_2(s,\bar t)$ for fixed $\bar t$ and the trivial value $\bar{x}_2(s,\bar t)=0$ is computed (Figure \ref{Fig:01}a) in order to demonstrate that there is a definite minimum during the evolution indicating initial convergence to the trivial solution followed by divergence beyond a critical point. This observation motivates our search for an appropriate definition of stability and critical point in the following section. Several shapes beyond that minimum are depicted in Figure \ref{Fig:01}b. To determine some characteristic features of these "postcritical curves", Figure \ref{Fig:01}c shows the normed deviations in $\bar x_2$.

\begin{figure}[!ht]
\centering
\includegraphics[width=0.95\textwidth]{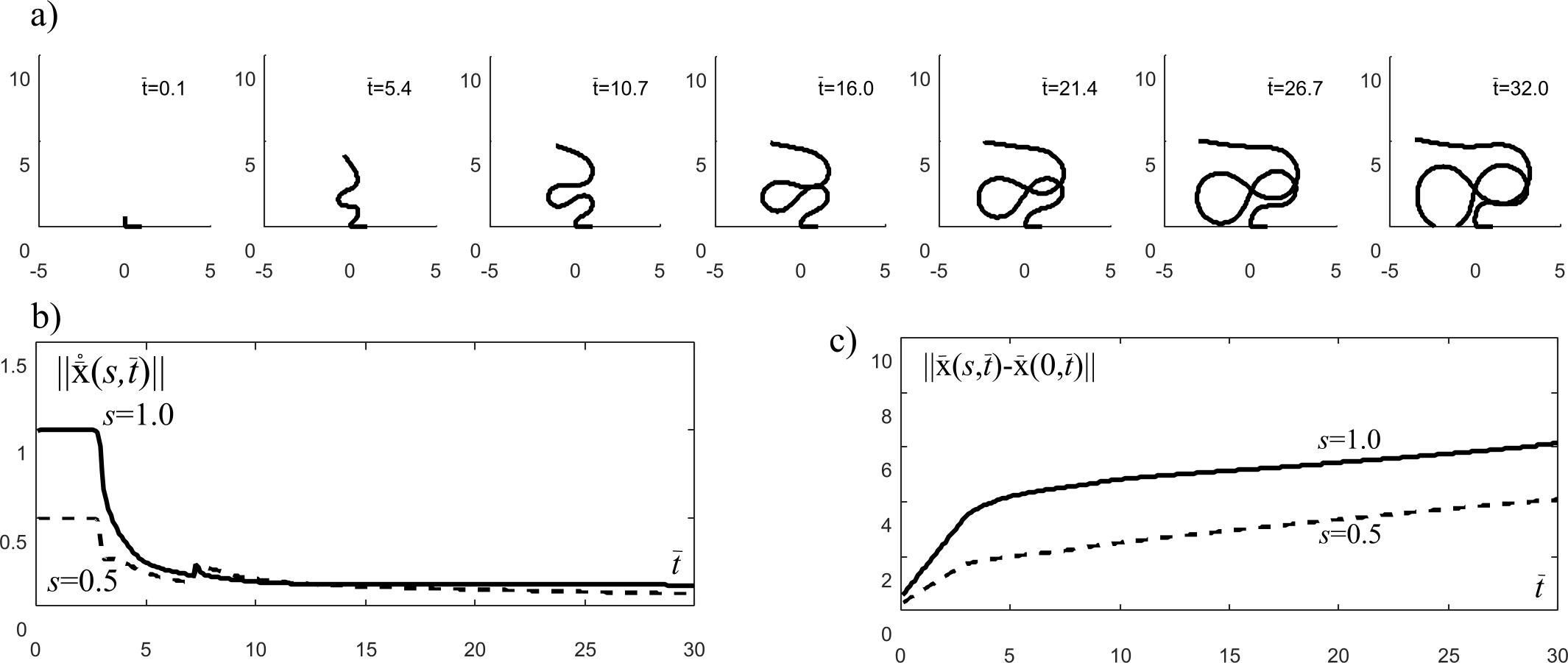}
\caption{Post-buckled shapes of the rod with $\bar{\mu}=1$ (pure Coulomb Friction) up to $t=32$ (a) Evolution of the shape. (b) The velocity of the free end ($s=1.0$) and the midpoint ($s=0.5$) of the curve. (c) The distance between endpoints and the distance between the midpoint and the fixed end.}
\label{Fig:02}
\end{figure}

Our results suggest that instead of a sudden loss of stability, the rod begins to become curved gradually. There is no well-defined post-critical shape. As Fig. 1(c) shows, the character of the "buckled shape" changes significantly even when the deviation from the trivial shape is very small, i.e. geometric nonlinearity is negligible. This is a natural consequence of considering a non-autonomous system, and it radically differs from classical problems of elastic stability. In particular the simulation results suggest that the rod has no inflexion points initially, and then the number of inflexions grows gradually. It is also interesting to note that for pure dry friction ($\bar \mu=1$) and large $\bar t$, the rod converges to a figure 8 (Figure \ref{Fig:02}(a) if we allow self-crossing of the rod.  In this case, the distance between the endpoints should theoretically remain bounded. The weakly increasing trend in Figure \ref{Fig:02}(b-c) is a consequence of using a regularized friction law (finite $\bar\beta$).  The existence of such a bound is important in various applications when the goal of the motion of the endpoint is to explore the environment (such as in the case of resource exploration during root growth). Clearly, adding self-contact to the model would change this picture, however such an addition is beyond the scope of this paper.

\subsubsection{Concentrated growth}
\begin{figure}[!ht]
\centering
\includegraphics[width=0.95\textwidth]{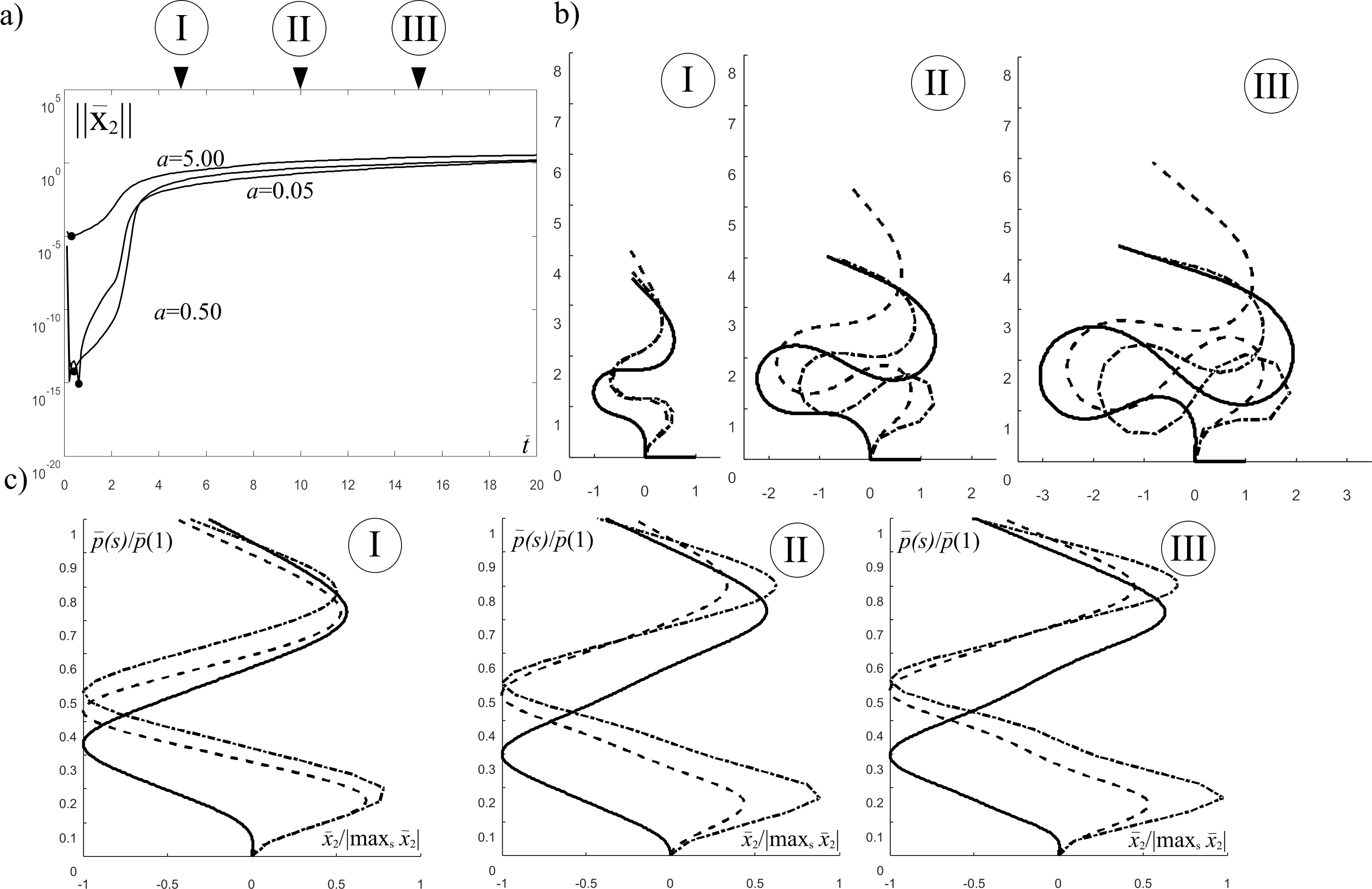}
\caption{From uniform to approximated concentrated growth under Coulomb friction ($\bar\mu=1.0$) at $a=5.00$ (solid line), $a=0.50$ (dash line) and $a=0.05$ (dash-dot line). (a) Norm of the difference between the computed and the trivial solutions in $x_2$. Black dot denotes the minimum of that difference. (b) Physical realization of the computed curves at $\bar{t}=5.0, 10.0, 15.0$. (c) Normed shapes for the same curves.}
\label{Fig:03}
\end{figure}

We now investigate the effect of concentrating the growth to the clamped end by applying eq. (\ref{eq:cgrowth}) at some distinct values of $a$. The value $a=5.00$ induces almost uniform growth, however $a=0.05$ heavily concentrates the growth to the vicinity of $s=0$. In this case the solution followed up to $\bar t=20$ (altogether 2000 timesteps). The computational results are summarized in Figure \ref{Fig:03}. We can draw similar conclusions as in the case of uniform growth but the post-critical shapes depend on the type of growth. Observe, that smaller value for $a$ leads to higher number in the inflexion points along the curve and that self intersection happens earlier for more concentrated growth.

\section{Stability of the trivial solution}
\label{Sec:stability}

Here we aim to investigate analytically the stability of the trivial solution of the growing rod problem. The numerical results in the previous section show that the rod converges to the trivial solutions initially, but as its length growth, this is replaced by a tendency to diverge from the trivial solution. Here we attempt to define and determine the critical length corresponding to destabilization. To this end, we develop an appropriate criterion of \textit{stability against infinitesimally small perturbations over a finite time interval}. As we have pointed out in the introduction, the existing definition of Finite-time stability is not appropriate to our purpose, since it depends on the response of the system to perturbations of a given finite size.

The model we have developed in Section \ref{Sect2} is a system of nonlinear partial differential equations. We use linearization for local analysis near the trivial solution (where $\bar x_2=0$). In order to avoid severe theoretical difficulties of the theory of infinite dimensional operators \citep{Zeidler}, we will perform stability analysis on a discretized, finite degree of freedom version of the model. In particular, we consider spatial discreatization via central finite differences. In the custom made MATLAB code an equidistant discretization of $[0,1]$ consisting of 1000 vertices is applied. In this way the discretized analogue of eq. (\ref{GE:linear}) takes the form of a linear, non-autonomous ODE
\begin{align}
\mathring y(\bar t) = A(\bar t)y(\bar t).
\label{eq:linearODE2}
\end{align}
Here, $y(\bar t)$ is a vector consisting of 1000 distinct values of $\bar x_2(.,\bar t)$. 

\begin{figure}[!ht]
\centering
\includegraphics[width=0.95\textwidth]{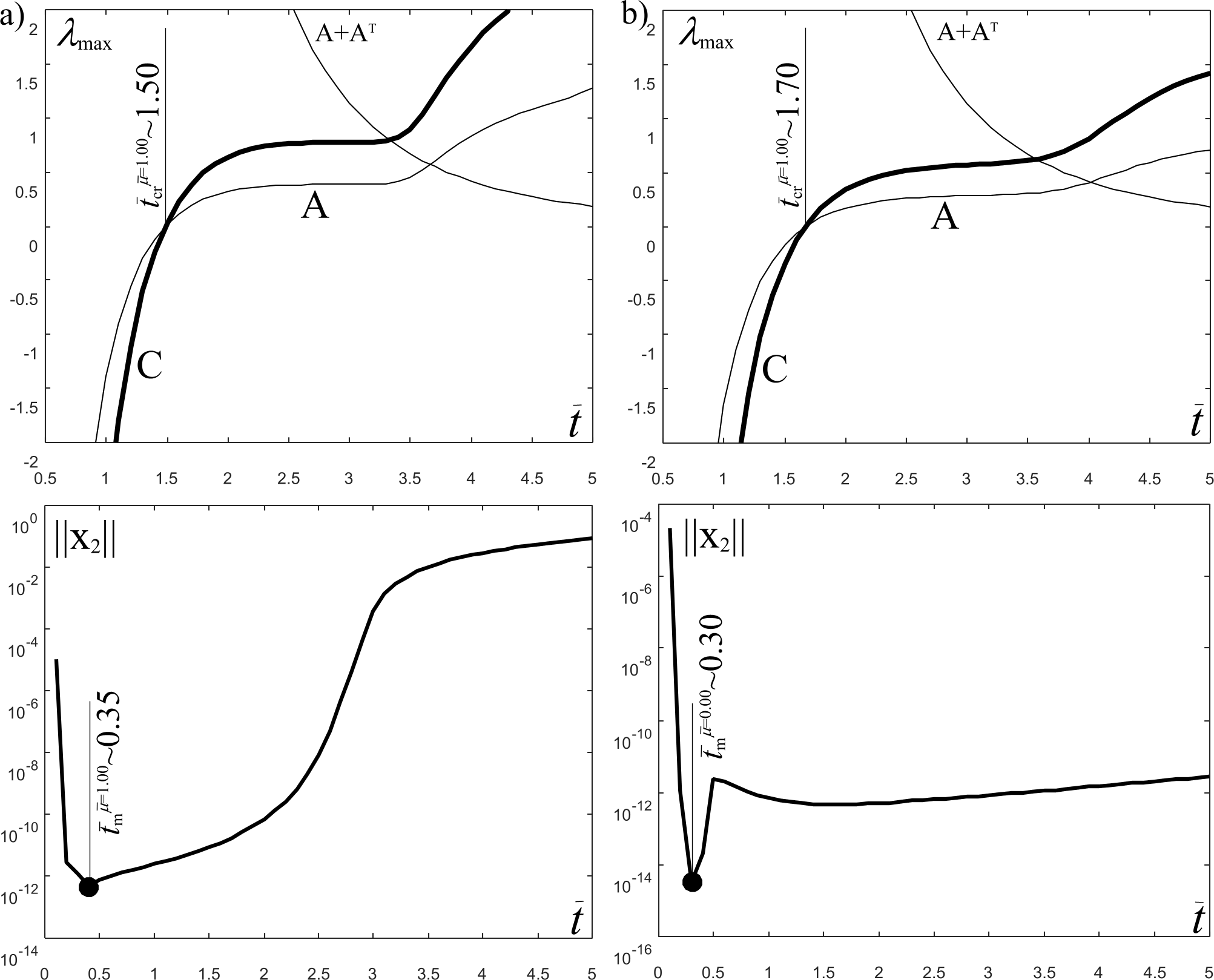}
\caption{The top panels show stability results for uniform growth (Figure \ref{Fig:01}) with (a) $\bar\mu=1$ and (b) $\bar\mu=0$.  The maximal eigenvalue, $\lambda_{max}$ is plotted for $A(t)$, $A(t)+A(t)^T$ and for $C(t)$. The bottom panels show the norm of $x_2$ as time is varied with circles denoting the minimum point of the $L_2$ norm. Observe, that the non-negativity of $A(\bar t)+A(\bar t)^T$ is well reflected in the sudden minimum in $\left\|x_2\right\|$.}
\label{Fig:05}
\end{figure}

As a starting point of the stability analysis, we revisit the classical notion of exponential stability of an autonomous linear vector-valued ordinary differential equation with an equilibrium solution at $y=0$:

\begin{definition} \label{def:expstab}
The ODE
\begin{align}
\mathring y(\bar t) = A y(\bar t)
\label{eq:linearODE}
\end{align}
 is exponentially stable if there exist positive scalars $c,\chi$ for which all solutions of the system satisfy for any $\bar t_1$ and $\bar t_2>\bar t_1$: 
\begin{align}
|y(\bar t_2)| \leq c|y(\bar t_1)|e^{-\chi (\bar t_2-\bar t_1)}
\label{eq:expconvergence}
\end{align}
with $|.|$ standing for an arbitrary norm.
\end{definition}
The conditions of exponential stability can be stated in several ways, as explained in standard textbooks. We use
\begin{theorem}[\cite{Chicone}, Theorem 2.60] \label{thm:stabilities} The following statements are equivalent
\begin{enumerate}
\item [(1)] Each eigenvalue of $A$ has negative real part.
\item [(2)] there exists a norm $|.|_*$ and a positive scalar $\chi$ for which solutions of the system satisfy for any $\bar t_1$ and $\bar t_2>\bar t_1$: 
\begin{align}
|y(\bar t_2)|_* \leq |y(\bar t_1)|_*e^{-\chi(\bar t_2-\bar t_1)}.
\label{eq:expconvergence2}
\end{align}
\item [(3)] the $y=0$ solution is exponentially stable.
\end{enumerate}
\end{theorem}
Unfortunately, our model \eqref{eq:linearODE2} is non-autonomous. The definition of exponential stability is applicable to non-autonomous systems as well, but it becomes a tough question how to test stability.  First, we tested the system matrix $A(\bar t)$ of the linearized (cf. eq. (\ref{GE:linear})) and discretized version of our problem with $\bar\mu=1.0$ and uniform growth for the eigenvalue property (1) of Theorem \ref{thm:stabilities} and found (Fig. \ref{Fig:05}) that all eigenvalues are negative if and only if \begin{align}
\bar{t} < \bar t_{cr}:=1.50.
\label{eq:tcr1}
\end{align}    
A similar critical value $\bar t_{cr}=1.70$ was found for viscous friction ($\bar{\mu}=0$).
At first glance, this result appears consistent with the observation that the rod tends to become straight during initial phases of motion, and diverges from the trivial solution later on. Unfortunately, it is well-known that when the matrix $A$ is time-dependent and non-symmetric, then satisfying property (1) of Theorem \ref{thm:stabilities} at all times in general does not imply stability \citep{josic2008unstable}. This is why we will use property (2), which is clearly a sufficient condition of exponential stability for non-autonomous systems.

The second challenge is how to define stability over a \emph{finite time-interval}. Clearly, the relation \eqref{eq:expconvergence} in the definition of exponential stability becomes useless, because any system free from singularities satisfies it over a finite interval if $c$ chosen large enough. That is why we propose is to \textit{define} stability by using property (2) of Theorem \ref{thm:stabilities} as follows:
\begin{definition}
Let $I\subset\xR$ be a closed interval. The non-autonomous linear ODE \eqref{eq:linearODE2}
is exponentially stable over $I$ if there exists a norm $|.|_*$ and a positive scalar $\chi$ such that any solution of the system satisfies for all $\bar t_1,\bar t_2\in I$, $\bar t_1<\bar t_2$ the inequality \eqref{eq:expconvergence2}.
\label{def:finitetimestab}
\end{definition} 
 
Clearly, this property cannot hold unless all eigenvalues of $A(\bar t)$ are negative for all $\bar t\in I$, i.e. unless $I\subset[0,\bar t_{cr})$ (closed at the left and open at the right end). In what follows, we demonstrate that the rod is exponentially stable over any closed sub-interval within  $[0,\bar t_{cr})$. 

If we choose the $L_2$ norm in \eqref{eq:expconvergence2}, then by using Gr\"onwall's inequality we arrive to the equivalent condition
\begin{align}
\frac{d}{d\bar t}|y(\bar t)|_{2}&=\\
\frac{d}{d\bar t}(y^T(\bar t)y(\bar t))^{1/2}&=\\
(y^T(\bar t)y(\bar t))^{-1/2}y^T(\bar t)A(\bar t)y(\bar t)&\leq -\chi|y(\bar t)|_{2},
\label{eq:dotxnorm}
\end{align}
which should hold for any vector $y\in \xR^n$. Note, that eq. (\ref{eq:dotxnorm}) is satisfied if and only if the symmetric matrix $A(\bar t)+A^T(\bar t)$ has negative eigenvalues for all $\bar t\in I$ \citep{johnson1970}. Throughout the paper, this property of a non-symmetric matrix is referred to as \emph{negative definiteness}. 

The system matrix of the rod problem has been tested numerically. The results suggest that it is not negative definite (Fig. \ref{Fig:05}). This means that despite the strictly negative eigenvalues of $A(\bar t)$ there are directions, in which the dynamics \eqref{eq:linearODE2} magnifies the lengths of vectors. Hence, using the $L_2$ norm in Definition \ref{def:finitetimestab} is not able to explain, why the rod initially converges to straight configurations. To obtain stronger results, we consider a set of norms of the form
\begin{align}
|y(\bar t)|_W:=\left |Wy(\bar t) \right|_{2},
\end{align}
where $W$ is an appropriately chosen square matrix. Then the condition \eqref{eq:dotxnorm} is replaced by 
\begin{align}
\frac{d}{d\bar t}{|y(\bar t)|_W}&=\\
\frac{d}{d\bar t}(y^T(\bar t)W^TWy(\bar t))^{1/2}&=\\
(y^T(\bar t)W^TWy(\bar t))^{-1/2}y^T(\bar t)W^TWA(\bar t)y(\bar t)&\leq -\chi|y(\bar t)|_{W}.
\label{eq:dotxnorm2}
\end{align}
Again, the inequality should hold for any $y\in \xR^n$, i.e. negative definiteness of the non-symmetric matrix $W^TWA(\bar t)$ is required. Accordingly, we need to verify the negative real part of all eigenvalues of $W^TWA(\bar t)+(W^TWA(\bar t))^T$ for all $\bar t\in I$. Assume that $A(\bar t)$ is diagonalizable and consider now the diagionalization of $A(\bar t)$ in the rod problem: 

\begin{align}
A(\bar t)=V(\bar t)^{-1}\Lambda(\bar t)V(\bar t),
\label{eq:decomp}
\end{align}

\noindent where $\Lambda(\bar t)$ is a diagonal matrix with the $\lambda_i(\bar t)$ eigenvalues ($i=1...n$) of $A(\bar t)$ in the main diagonal, and the row vectors of $V(\bar t)$ are the corresponding left eigenvectors. In the following we choose $W=V(\bar t)$.

\begin{lemma}
Let $W$ denote the (square matrix) of left eigenvectors of $A(\bar t)$ for some $\bar t$ as above. If the nonsingular matrix $A(\bar t)$ has negative eigenvalues then $B(\bar t):=W^TWA(\bar t)$ is negative definite. 
\end{lemma}

Proof: Let $\left\langle.,.\right\rangle$ denote the scalar product in $\xR^n$. Negative definiteness of $B(\bar t)$ means $\left\langle y,W^TWA(\bar t)y\right\rangle<0$ for all $y\in\xR^n\setminus\{0\}$. Our assumptions above yield that $W$ is regular, hence
\begin{align}
\left\langle y,W^TWA(\bar t)y \right\rangle&=\\
\left\langle Wy,WW^{-1}\Lambda(\bar t)Wy\right\rangle&=\\
\left\langle Wy,\Lambda(\bar t)Wy\right\rangle &=\\
\sum_{i=1}^{n}\lambda_i(\bar t)(Wy)_i^2&<0,
\end{align}
where $(Wy)_i$ is the $i$th element of the vector $Wy$.
\begin{corollary}
If $A(\bar t)$ is singular with all nonzero eigenvalues being negative, then $B(\bar t)$ is negative semi-definite.
\end{corollary}
\begin{corollary}
If $V(\bar t)=V$ in eq. (\ref{eq:decomp}) is time independent, then the choice $W=V$ makes $B(\bar t)=W^TWA(\bar t)$ negative definite for all $\bar t\in[0,\bar t_{cr})$, hence, stability is assured.
\end{corollary}

Corollary 2 suggests, that the stability of systems keeping the eigenspace of $A(\bar t)$ fixed during the evolution can be simply verified by monitoring the eigenvalues of $A(\bar t)$. However, time invariance of $V(\bar t)$ is not granted in many problems (including our rod problem), hence the choice of $W$ is rather arbitrary. In our work we choose $W=V(\bar t_{cr})$, which - following Corollary 1 renders $B(\bar t_{cr})$ to a negative semi-definite matrix. Then the maximal eigenvalue of the symmetric matrix $C(\bar t):=B(\bar t)+B(\bar t)^T=W^TWA(\bar t)+(W^TWA(\bar t))^T$ can be investigated numerically for $0\leq \bar t \leq \bar t_{cr}$ (Fig. \ref{Fig:05}). 
The analysis reveals that all eigenvalues of $C(\bar t)$ are negative for all values of $\bar t\in(0,\bar t_{cr})$. This brings us to the final conclusion that the trivial shape of the growing rod is exponentially stable over any closed time interval within $[0,\bar t_{cr})$ but not stable for any time interval including or extending beyond $\bar t_{cr}$. This finding explains the results of the numerical simulation presented in Sec. 3. 

We present the stability results for uniform growth in Figure \ref{Fig:05}. Note, that the region of exponential stability \emph{significantly} extends beyond the minimum of the $L_2$ norm of the solution. Hence, there is a significant amount of time, when the $L_2$ norm of solutions may grow, nevertheless there exists another norm, which provably decreases for any solution.

\section{Experimental verification of post-critical behavior}

\begin{figure}[!ht]
\centering
\includegraphics[width=0.95\textwidth]{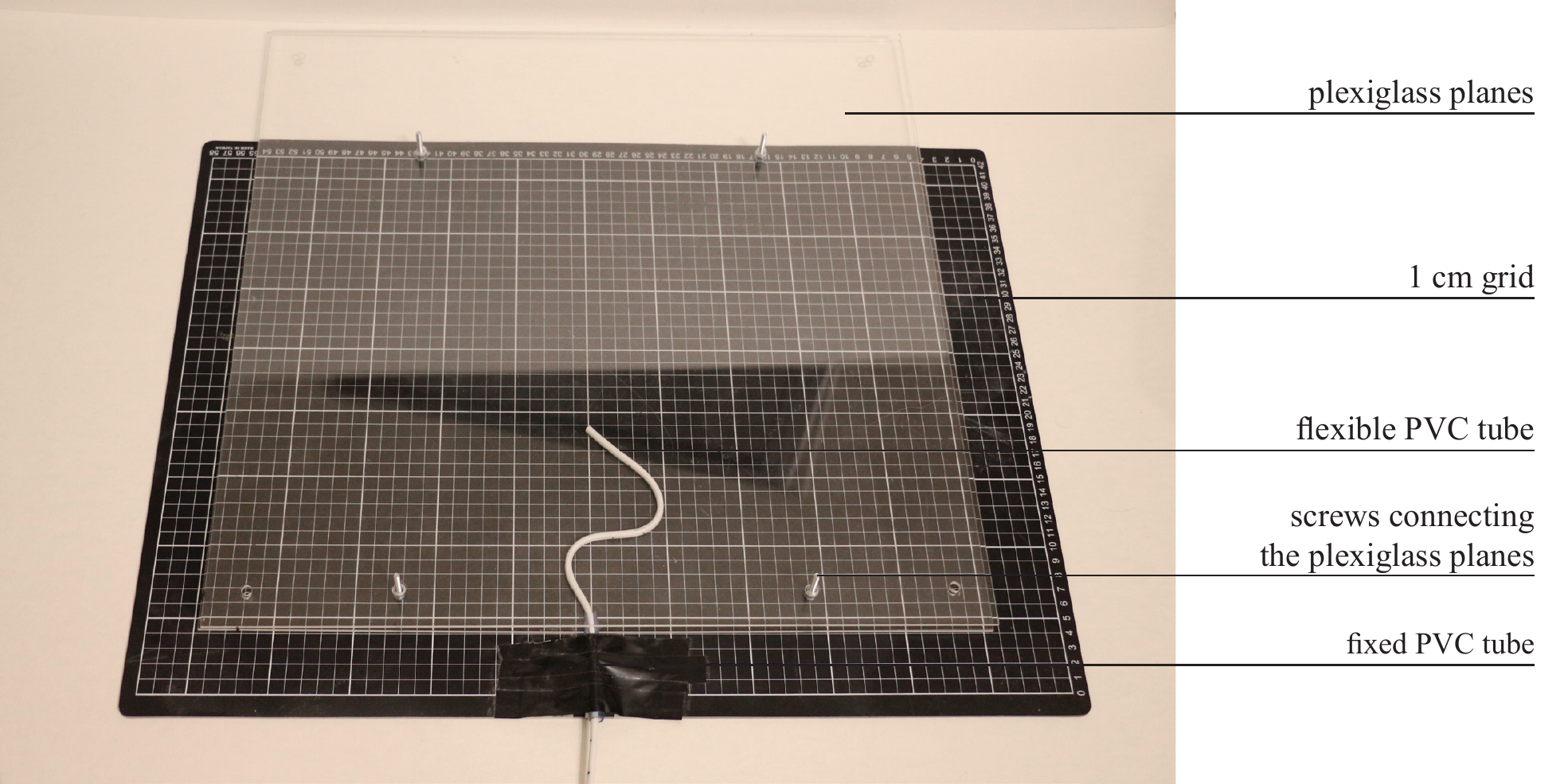}
\caption{The experimental setup.}
\label{fig:experiment}
\end{figure}

In order to verify our numerical results, a series of physical experiments has been conducted. A PVC electric cable insulation tube of nominal diameter 5 mm and wall of thickness 0.5mm has been driven through a fixed steel tube and pushed slowly into a thin gap between two horizontal plexiglass plates (Fig. \ref{fig:experiment}). The gap was set to be slightly narrower than the diameter of the tube thus the motion of the cable was accompanied by significant amount of sliding friction.  This setup mimics growth of a clamped-free rod concentrated at the clamped end. Each experiment was ended when the tube established self-contact.

The model parameters of the experimental setup were estimated based on force and length measurements as described in the Appendix. The frictional force was approximated by a force proportional to cable length (which is consistent with the dry friction law used in the model) with an approximate value of $\mu=20.03$N/m. Despite geometric imperfections of the plexiglass plate and deflection due to its own weight, we found that this value was not strongly affected by the exact placement of the tube within the device. We believe that this beneficial behavior is due to the compliance of the thin-walled cross-section of the tube. The bending stiffness was calculated as the product of the estimated Young's modulus ($Y$) and the moment of inertia ($I$). $I$ was determined from measurements of the tube diameter (average: $5.04$mm) and wall thickness (average: $0.53$mm). The modulus of elasticity ($Y=22.90$N/mm$^2$) was estimated from experimental force-displacement curves of the tube under tension. These curves also revealed that a linear elastic model is a reasonable approximation.

\begin{figure}[!ht]
\centering
\includegraphics[width=0.95\textwidth]{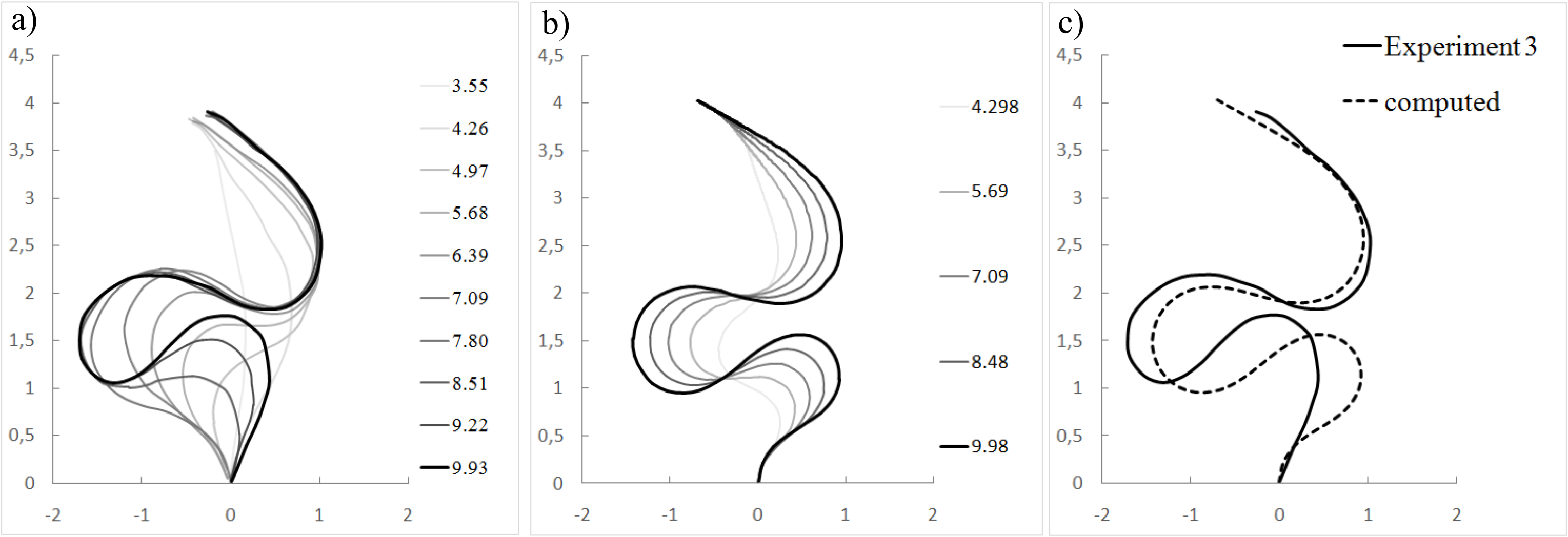}
\caption{Nondimensionalized comparison of one of the experimental results (a) with simulation (b). Shapes are identified by their arclength $\bar p(1,\bar t)$. (c) shapes at $\bar p(1,\bar t)=9.93$, shortly before self-contact.}
\label{fig:expresults}
\end{figure}

\begin{figure}[!ht]
\centering
\includegraphics[width=0.7\textwidth]{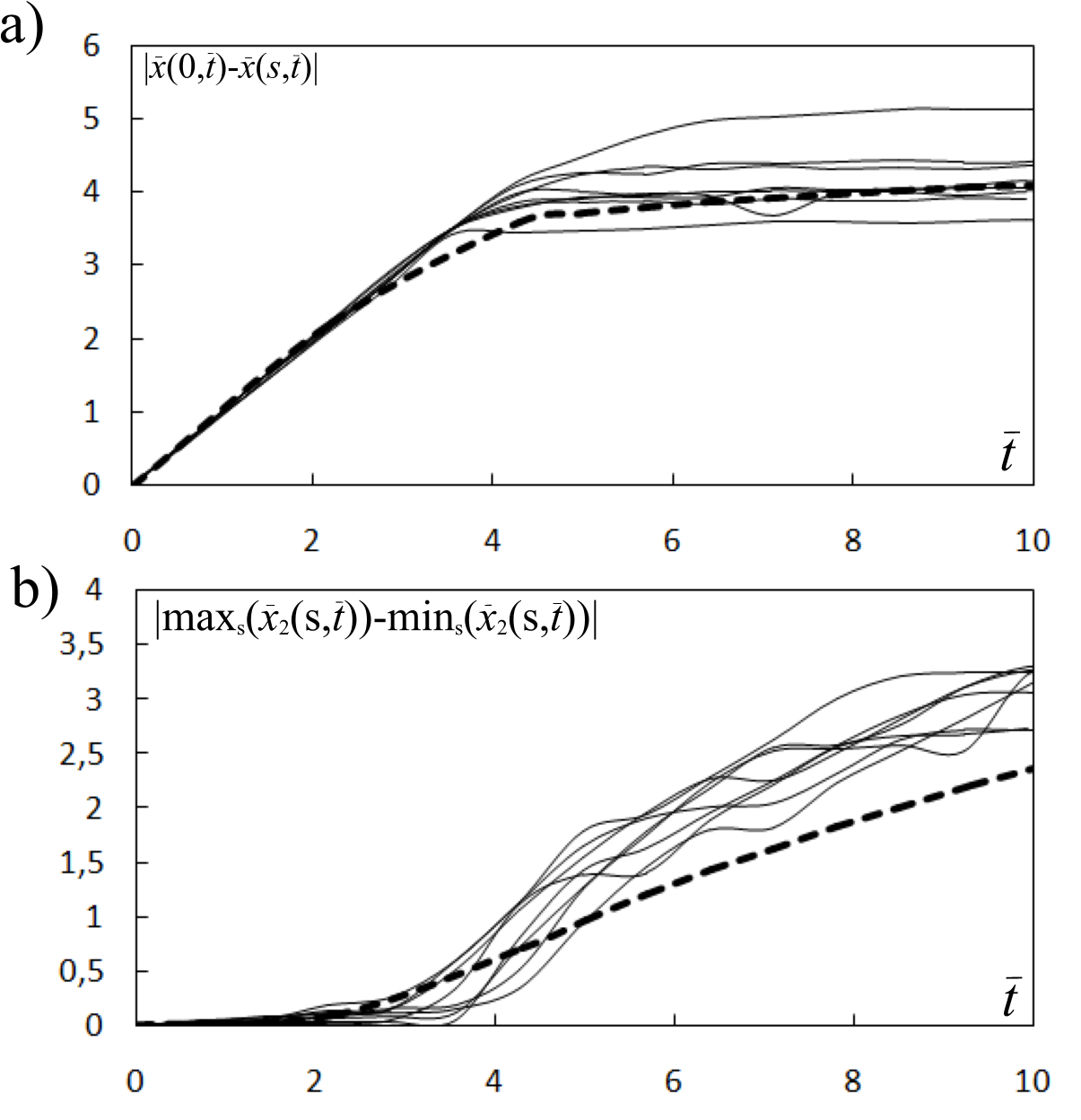}
\caption{Distance between the endpoints (a) and the width of the rod (b) as functions of $\bar t$ in all experiments (thin solid lines) and simulations (dashed lines).}
\label{fig:expmeasures}
\end{figure}

Photographs of the quasi-static experimental motion were taken and the shape of the tube was reconstructed with the help of WebPlotDigitizer \citep{WPD}. By using the estimated values of $\mu$ and $YI$, we rescaled the numerical results to make them directly comparable with simulation results of the non-dimensionalized model. In particular, the experimentally obtained physical lengths were multiplied by the factor $(YI)^{-1/3}N^{1/3}$=$(YI)^{-1/3}\mu^{1/3}=0.3547$m$^{-1}$ in accordance with \eqref{eq:lengthscaling}. Figure \ref{fig:expresults} shows the evolution of rod shape in the experiments (panel a) and according to the simulation (b). Shortly before self contact the experimental and computed shapes are plotted in panel (c). We see that there is a high degree of similarity between the two shapes. To make this observation more precise we show the evolution of the non-dimensionalized distance between the endpoints of the rod ($|\bar x(0,\bar t)-\bar x(1,\bar t)|$) as well as the 'width' of the rod (i.e. $\left|\max_s(\bar x_2(s,\bar t))-\min_s(\bar x_2(s,\bar t))\right|$) as functions of $\bar t$ (or equivalently non-dimensional arc-length) for all experiments, together with the corresponding simulation results (Fig. \ref{fig:expmeasures}).

\section{Conclusions}
\label{Sect6}

We examined in this paper the growth of an elastica in contact with a frictional surface. A mechanical model of the problem has been formulated, which took the form of a non-conventional elastic stability problem. We found that the growing rod initially converges to a trivial (straight) configuration, but later it tends to diverge from it. Identifying a critical point associated with the loss of stability required a new concept of stability against infinitesimal perturbations over a finite time interval. After developing an appropriate definition (motivated by the classical notion of exponential stability), we were able to determine the critical point by using a combination of analytic tools and numeric eigenvalue analysis. The post-critical behavior of the rod was then studied numerically, using a custom-made solver based on an extended energy approach and variational principles. We highlighted fundamental differences between this phenomenon and classical buckling, including gradual divergence instead of sudden transition, and the lack of a well-defined post-critical shape. The asymptotic behavior of the rod for large lengths was studied numerically. We showed that in the case of dry friction the rod develops into a figure 8 pattern. 

The problem studied here has many potential applications. The growth of plant roots in biology \citep{whiteley1982buckling,mathur2001cell,bengough2011root,datta2011root} is being studied intensively with focus on underlying biochemical mechanisms. Nevertheless the buckling of roots, which tends to occur when a root tip penetrates into a hard layer of soil, has also been studied \citep{clark2003roots,bengough2005root,silverberg20123d}. We believe that future work based on our modeling approach will provide useful insights into the connection between mechanical constraints of growths and biochemical mechanisms. Another potential application is the design of expanding soft manipulators in robotics \citep{hawkes2017soft}. While some types of manipulator design eliminate sliding-induced friction, others need to cope with it, and thus understanding the mechanical limitations of light-weight, and slender designs is of particular interest.

From the point of view of mechanical modeling, growth is often equivalent of having non-stationary contact constraints. For example, concentrated growth at a fixed endpoint is analogous to pushing a flexible rod of constant length through a stationary environment. We not only exploited this analogy in our experiments, but also demonstrated thereby the relevance of our approach to various applications such as feeding cables into conduits or stents into veins  \citep{dunn2007macroscopic,vad2010determination}. Mechanical models of animal whiskers and other tactile sensors \citep{goss2016loading,cutkosky2016force} sliding against a fixed surface also fit within the framework of our study after minor extensions of the modeling framework.

\section*{Acknowledgment}
The paper was supported by the J\'anos Bolyai Research Scholarship of the Hungarian Academy of Sciences [SA], by Grant 104501 of the National Research, Development, and Innovation Office, Hungary [VP, SA] and by the \'UNKP-17-4-III New National Excellence Program of the Ministry of Human Capacities, Hungary [VP].




\appendix
\section{Estimation of model parameters for the experimental setup}

All experimental results, both material and shape measurements, are available in the supplementary material.

\subsection{Dimensions of cross section}
The diameter $h$ and the wall thickness $w$ have been measured at 20 different cross-sections, yielding $h=5.04\pm0.027 mm$, $w=0.53\pm0.039mm$ (first value is the average, and the second is the  standard deviation). These values were used to estimate the area of the cross-section ($A=7.53mm^2$) and its moment of inertia ($I=19.61mm^4$).

\subsection{Frictional force}
10$+$10 measurements have been done in two orthogonal directions. A tube of length $50$cm was placed between the plates with a straight initial configuration and pulled out manually at a rate of $1$cm$/$s. The pulling force was measured by an analogue force meter at various lengths of the tube in steps of $5$cm. The average force $F$ and its standard deviations are depicted in Fig. \ref{fig:measurements}(a) as functions of tube length $l$. Though the force-length curve is significantly nonlinear, it has been approximated by a straight line $F=\mu l$ using linear regression, yielding an estimated value $\mu=20.03 N/m$ of the frictional force.

\subsection{Modulus of elasticity}
\begin{figure}[!ht]
\centering
\includegraphics[width=0.80\textwidth]{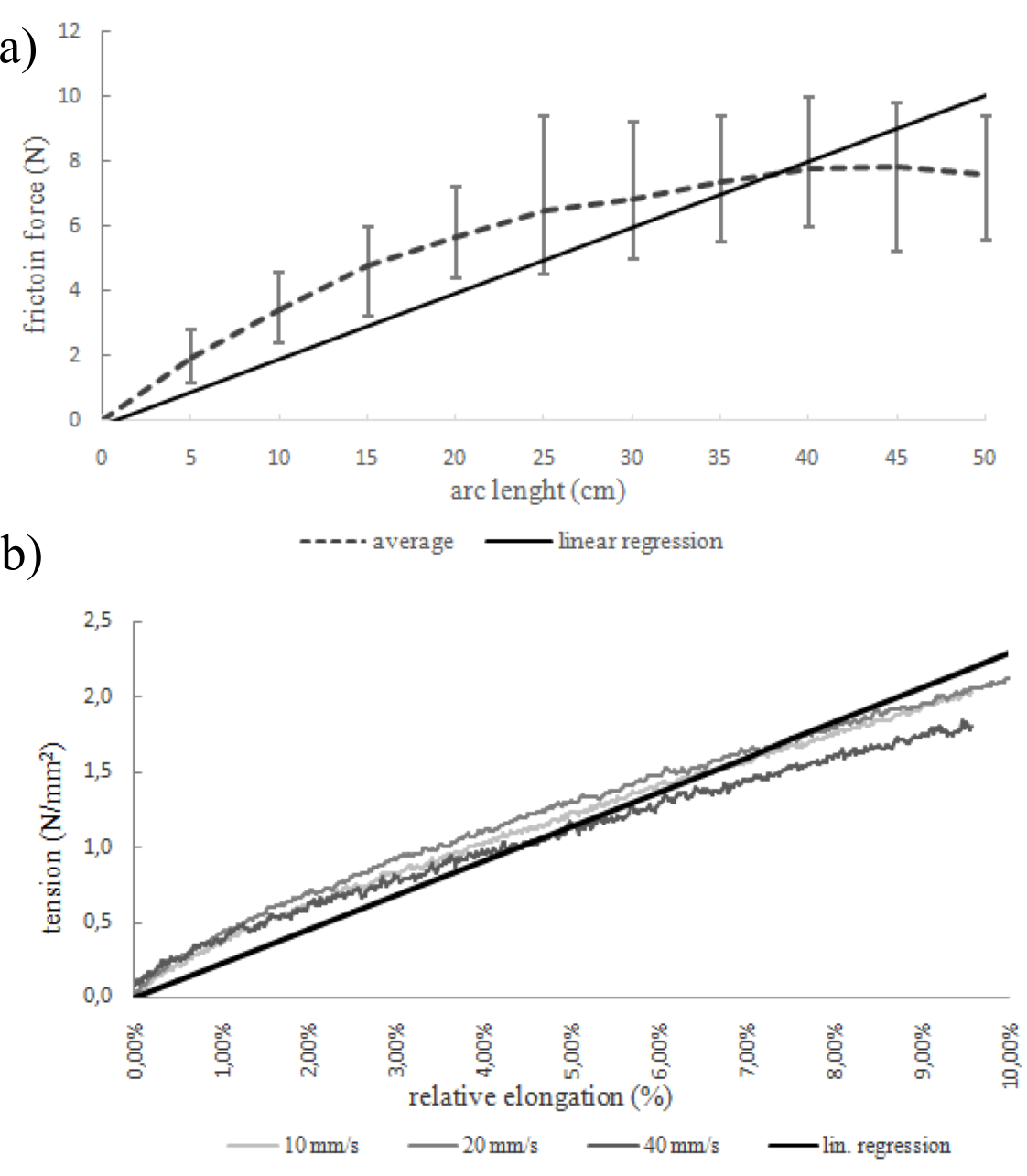}
\caption{Friction force as function of length (a) and averages of three stress-strain diagrams of the tube at various rates of elongation (b). }
\label{fig:measurements}
\end{figure}

3 times 3 force displacement curves of  specimens of length $l=67.0 mm$ have been recorded by a ZWICK Z150 testing machine at pulling velocities of $10$, $20$, and $40mm/s$. Each experiment was finished when the elongation was $30 mm$. The diagrams (Fig. \ref{fig:measurements}(b)) were nearly linear and only slightly velocity-dependent. By fitting a line to the average of all 9 measurements, we obtained the estimate of $Y=22.90$N/mm$^2$.

\bibliographystyle{elsarticle-harv}
\bibliography{References}




\end{document}